\documentclass[showpacs,prb,reprint,preprintnumbers,amsmath,amssymb,superscriptaddress,aps]{revtex4-1}
\usepackage{graphicx}
\usepackage{dcolumn}
\usepackage{bm}

\begin{document}
\title{Carrier-induced refractive index change and optical absorption\\ 
in wurtzite I\lowercase{n}N and G\lowercase{a}N: Fullband approach}

\author{Ceyhun \surname{Bulutay}}
\email{bulutay@fen.bilkent.edu.tr}
\author{Cem Murat \surname{Turgut}}
\affiliation{Department of Physics, Bilkent University, Ankara 06800, Turkey}
\author{N. A. \surname{Zakhleniuk}}
\affiliation{School of Computer Science and Electronic Engineering, 
University of Essex, Wivenhoe Park, Colchester, CO4 3SQ, United Kingdom}
\date{\today}

\begin{abstract}
Based on the full band electronic structure calculations,  first 
we consider the effect of $n$-type doping 
on the optical absorption and the refractive index in wurtzite InN and GaN. 
We identify quite different dielectric response in either case; while 
InN shows a significant shift in the absorption edge due to $n$-type doping, 
this is masked for GaN due to efficient cancellation of the Burstein-Moss 
effect by the band gap renormalization. Moreover, for high doping levels 
the intraband absorption becomes significant in InN. For energies below 
1~eV, the corresponding shifts in the real parts of the dielectric function 
for InN and GaN are in opposite directions. Furthermore, we observe that 
the free-carrier plasma contribution to refractive index change becomes 
more important than both band filling and the band gap renormalization for 
electron densities above 10$^{19}$~cm$^{-3}$ in GaN, and 10$^{20}$~cm$^{-3}$ in InN. 
As a result of the two different characteristics mentioned above, the 
overall change in the refractive index due to $n$-type doping is much 
higher in InN compared to GaN, which in the former exceeds 4\% for a 
doping of 10$^{19}$~cm$^{-3}$ at 1.55~$\mu$m wavelength. 
Finally, we consider intrinsic InN under strong photoexcitation which 
introduces equal density of electron and holes thermalized to their 
respective band edges. The change in the refractive index at 1.55~$\mu$m is 
observed to be similar to the $n$-doped case up to a carrier density 
of 10$^{20}$~cm$^{-3}$. However, in the photoexcited case this is now 
accompanied by a strong absorption in this wavelength region due 
to $\Gamma^v_5 \to \Gamma^v_6$ intravalence band transition. 
Our findings suggest that the alloy composition of In$_x$Ga$_{1-x}$N can be 
optimized in the indium-rich region so as to benefit from high carrier-induced 
refractive index change while operating in the transparency region to 
minimize the losses. These can 
have direct implications for InN-containing optical phase modulators and lasers.
\end{abstract}

\maketitle

\section{Introduction}
The research efforts on InN and In-rich InGaN have been intensifying worldwide. As well as their 
traditional applications on solid-state lightning and lasers, new possibilities begin to 
flourish such as photovoltaics and chemical sensing.\cite{wu09} From the electronic structure 
point of view, InN has a number of unique properties.\cite{rinke} Its band gap for the wurtzite phase is 
commensurate with the 1.55~$\mu$m wavelength for fiber optics when it
forms an alloy with a small amount of GaN.
Furthermore, the small conduction band effective mass enables interesting band-filling effects.
Even though the latter giving rise to Burstein-Moss effect\cite{burstein54,moss54} was the center 
of focus in the recent re-assessment of the high quality InN samples,\cite{wu02a,davydov,wu02b,wu04} 
its device implications have not been given the full attention it deserved. Among the viable 
applications are the carrier-induced optical phase modulators,\cite{vinchant} 
and tunable Bragg reflector and filters.\cite{deppe,weber} These devices are expected to play 
essential role, for instance, for the signal processing directly in the optical domain in the future 
high bit rate optical communication networks. 

It was known from the early days of edge-emitting semiconductor lasers that carrier-induced 
refractive change affects the optical beam quality,\cite{kirkby77} such as mode guiding 
along the junction plane, self-focusing, filament formation, frequency chirping under 
direct modulation.\cite{olsson81,dutta84} The refractive index tunability by carrier 
injection was studied in the previous decade for the Indium-Group-V semiconductor 
compounds and alloys other than InN.\cite{bennett,chusseau,paskov97}  
In the case of InGaN laser structures, there have been recent theoretical\cite{chow00} and 
experimental\cite{schwarz03,rowe03} studies on the carrier-induced refractive index change and 
the linewidth enhancement factor. However, these studies considered only Ga-rich limit of InGaN alloys.

The purpose of this work is to offer a comparative theoretical account of carrier-induced 
refractive index change in both InN and GaN, so as to form a basis for the In-rich InGaN alloys.
Primary attention is given to $n$-type doping which is the prevalent type among the routinely 
grown samples.\cite{wu09} For the case of InN, the photoexcited bipolar 
carrier-induced dielectric effects are also examined. The crux of our study is based on the 
Bennett-Soref-Del Alamo approach\cite{bennett} which includes band-filling, 
band gap renormalization (BGR) 
and plasma contributions. However, rather than a model band structure, we employ a rigorous full band 
technique to represent the band filling; preliminary account of this work was given 
in Ref.~\onlinecite{pssc08}.
In Sec.~II we present the theoretical details, this is followed by our results in Sec.~III. 
A self-critique of our model and further discussions are 
provided in Sec.~IV, ending with our conclusions in Sec.~V. The Appendix contains our 
empirical pseudopotential parameters used for wurtzite InN.

\section{Theory}
\begin{figure}[h]
\begin{center}
\includegraphics[width=9cm]{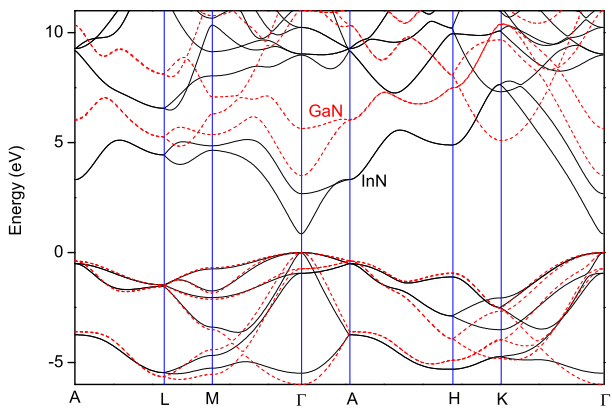}
\caption{\label{fig1} The empirical pseudopotential band structures 
for InN (solid) and GaN (dashed) used in this work.}
\end{center}
\end{figure}

First, we would like to describe the electronic structure on which our computations are based.
In the case of GaN, we use the local empirical pseudopotential method that we fitted to the existing 
experimental data, putting special emphasis on the conduction band behavior.\cite{prb00}
For InN, we use the nonlocal empirical pseudopotential method, treating the angular-momentum 
channels $p$ and $d$ for In, and $p$ for N as nonlocal.\cite{sm04} Further details are given in the 
Appendix section. The corresponding band gap for InN comes out as 0.85~eV which agrees very well with 
the recent first principles results\cite{carrier05,bagayoko05} but it is higher than the 0.64~eV value established 
experimentally.\cite{wu09} Figure \ref{fig1} shows the band structures of wurtzite InN and GaN.

\begin{figure}[h]
\begin{center}
\includegraphics[width=8.5cm]{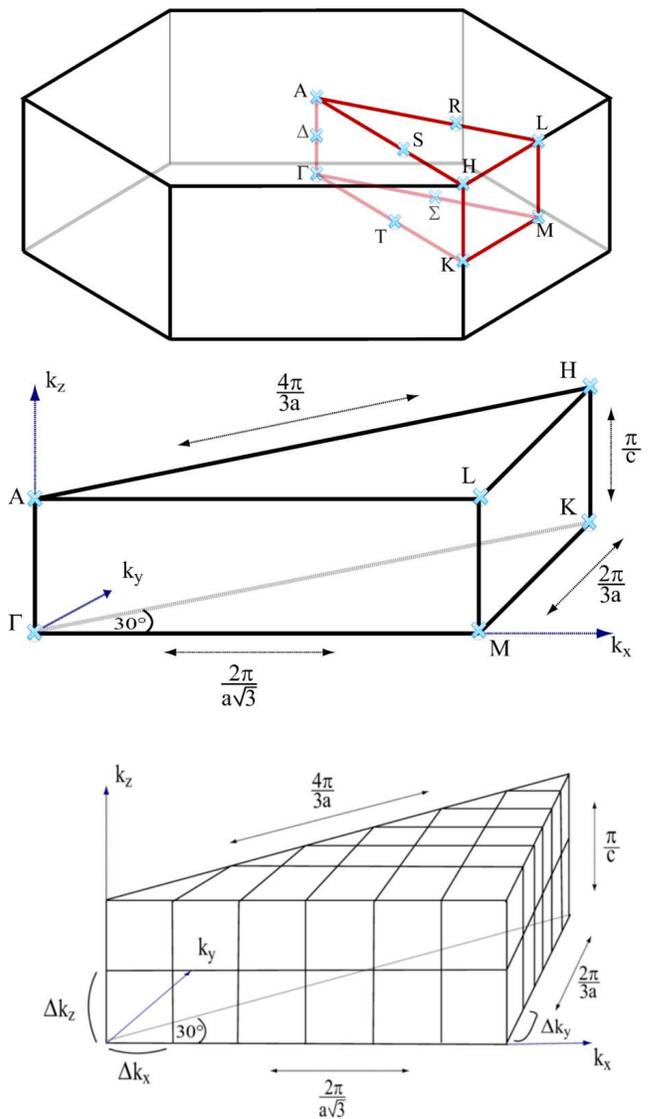}
\caption{\label{fig2}Schematic (not to scale) of the Brillouin 
zone of the wurtzite structure (top), its irreducible 
Brillouin zone (middle), and its tessellation (bottom); 
each volume is further divided into tetrahedra (not shown). 
$a$ and $c$ are the lattice constants in the basal plane and 
perpendicular directions, respectively.}
\end{center}
\end{figure}

As the Burstein-Moss effect\cite{burstein54,moss54} primarily shifts the absorption edge due 
to band filling, the ultimate 
quantity to be computed is the imaginary part of the optical (i.e., the long wavelength limit) dielectric 
tensor which is given in Gaussian units by\cite{hughes}
\begin{eqnarray}
\label{im_eps}
\mbox{Im}\left\{ \epsilon^{ab}(\omega)\right\} & = & \frac{e^2}{\pi}\sum_{v,c}
\int_{{\mbox{\begin{scriptsize}{BZ}\end{scriptsize}}}} d\textbf{k}\,
r^a_{vc}(\textbf{k})r^b_{cv}(\textbf{k})
\nonumber \\ & & \times
\delta\left( E_c(\textbf{k})-E_v(\textbf{k})-\hbar\omega\right)\, ,
\end{eqnarray}
where $a, b$ are the Cartesian indices. $r^a_{vc}(\textbf{k})=p^a_{vc}(\textbf{k})/(im_0\/ 
\omega_{vc}(\textbf{k}))$, where
 $p^a_{vc}(\textbf{k})$ is the momentum matrix element, $m_0$ is the free electron mass, 
 $\omega_{vc}(\textbf{k})\equiv \omega_v(\textbf{k})-
\omega_c(\textbf{k})$, where $\hbar\omega_n(\textbf{k})\equiv E_n(\textbf{k})$ is the 
energy of the band $n$, at the wave vector \textbf{k}.
The specific band label $v$ ($c$) represents filled (unfilled) bands, where for the clean 
cold semiconductors, the band occupation factors 
are taken as either one or zero. The volume integration in Eq.~(\ref{im_eps}) is over 
the first Brillouin zone (BZ) which can be reduced to the irreducible BZ using the 
symmetry relations of the BZ (see, Fig.~\ref{fig2}).

\begin{figure}[h]
\begin{center}
\includegraphics[width=9cm]{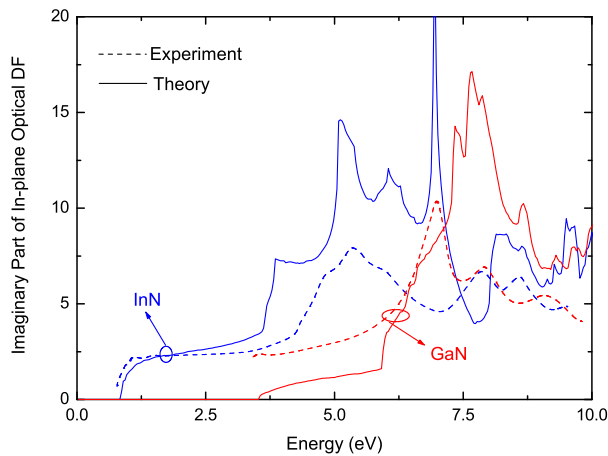}
\caption{\label{fig3} Imaginary part of the in-plane optical dielectric function 
for the intrinsic InN (blue) and GaN (red), comparing our theoretical (solid) results 
with the experimental (dashed) data of Refs.~\onlinecite{goldhahn} and \onlinecite{benedict}.}
\end{center}
\end{figure}

The Dirac delta term in Eq.~(\ref{im_eps}) automatically reduces the volume integration 
to a surface formed by the \textbf{k} points which allow direct (vertical) 
transitions from a filled state to an unfilled state with an energy difference 
corresponding to the chosen photon energy. Such surface integrations routinely 
appear in the density of states, effective mass and response function 
calculations.\cite{martin-book} They can be efficiently calculated numerically 
using the Lehmann-Taut method.\cite{lehmann}
This technique has also been shown to work for the second-order 
nonlinear optical response function\cite{moss87} and also for the full band 
phonon scattering rate calculation.\cite{prb00} In this work, for the required 
accuracy, we divide the irreducible BZ into a mesh of 
$40\times 40 \times 40$ along the basal plane and the $c$-axis directions 
(see, Fig.~\ref{fig2}). In Fig.~\ref{fig3}, we show the imaginary part of the 
optical dielectric function for the intrinsic (i.e., undoped) InN and GaN.
The ordinary tensor component which corresponds to the electric field 
parallel to basal plane of the hexagonal crystal (i.e., in-plane) 
polarization is shown. For GaN, we observe a discrepancy between the theoretical 
and experimental\cite{benedict} values close to the band edge. On the other 
hand, our theoretical result for InN matches well with the 
rigorous first-principles studies,\cite{furthmuller,jin07} and also in reasonable 
agreement with the experiment.\cite{goldhahn}

\begin{figure}[h]
\begin{center}
\includegraphics[width=9cm]{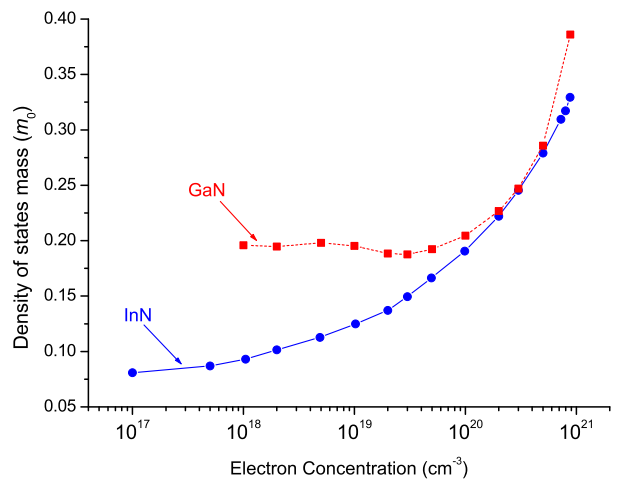}
\caption{\label{fig4} The density-of-states effective mass variation as a function of band-filling
for the conduction band electrons in InN and GaN. The latter is computed in this work, whereas 
the former is taken from Ref.~\onlinecite{carrier05}.}
\end{center}
\end{figure}

Another physical mechanism that accompanies band-filling is the BGR 
due to many-body interactions among the carriers.\cite{berggren} In highly $n$-doped 
InN samples it is observed that this partially cancels the Burstein-Moss effect.\cite{wu02b} 
The electron-electron and electron-ion contributions within the random 
phase approximation are given in Gaussian units by\cite{berggren,wu02b}
\begin{eqnarray}
\label{bgr}
\Delta E_{e-e} & = & -\frac{2e^2k_F}{\pi\epsilon_s}-
\frac{e^2k_{TF}}{2\epsilon_s}\left[ 1-\frac{4}{\pi}\arctan\left(
\frac{k_F}{k_{TF}} \right) \right] \\
\Delta E_{e-i} & = & -\frac{4\pi e^2 N}{\epsilon_s a^*_B k^3_{TF}}\, , 
\end{eqnarray}
where $N$ is the electron (ion) density and $\epsilon_s$ is the static 
permittivity for which we use the experimental value of 9.5 (15.3) for 
GaN (InN).\cite{azuhata,zubrilov}
$k_{F}=(3\pi^2N)^{1/3}$ and  $k_{TF}=2\sqrt{k_F/(\pi a^*_B)}$ are, 
respectively, the Fermi and Thomas-Fermi wave numbers, 
$a^*_B=0.529\, \epsilon_s m_0/m^*$ is the effective Bohr radius in \AA, $m^*$ 
is the conduction band effective mass. The latter quantity rapidly deviates 
with increase of energy from its band edge value, especially in the case of InN.
Therefore, in these expressions we use an energy-dependent density of states 
effective mass value corresponding to the level of band-filling.
Figure~\ref{fig4} shows for the conduction band these values used in this work  
for InN and GaN. The curve for InN is taken from the first-principles 
results.\cite{carrier05} For GaN, it has been computed using the 
same Lehmann-Taut method, but with a finer grid than used above, such 
as $60\times 60 \times 60$ or more.
In the case of photoexcitation which we consider for InN, actually we have a 
two-component quantum system (i.e., electron and hole gases). In the interests of simplicity,
we adapt Eq.~(\ref{bgr}) by replacing the electron effective mass with the 
reduced electron-hole effective mass, $\mu_{eh}$ given by 
$\mu_{eh}^{-1}=m^{-1}_e+m^{-1}_h$ where 
$m_e$ and $m_h$ are the energy-dependent electron and hole effective masses, 
respectively. The latter is almost constant for the densities dealt in this 
work and therefore is taken to be 0.84$m_0$ for InN.

Having determined the imaginary dielectric function of a semiconductor with partially filled band(s), 
subject to BGR, the next aim is to compute their effect on the refractive index. 
Their interrelation is established by the causality principle which has to 
be satisfied by any physical response function.\cite{klingshirn} In mathematical terms, this is governed 
by the Kramers-Kronig relation; for the real part, the corresponding expression, 
suppressing the Cartesian indices is given by
\begin{equation}
\label{kkr}
\mbox{Re}\left\{ \epsilon(\omega)\right\}=1+\frac{2}{\pi}\mathcal{P} \int_{0}^{\infty} \frac{\omega'
\mbox{Im}\left\{\epsilon(\omega')\right\}}{\omega'^2-\omega^2} d\omega' \, .
\end{equation}
Computationally, it requires the knowledge of the imaginary part of the dielectric function for all 
energies; we have verified the convergence of our results by including 60 bands and 
accounting for the higher energy behavior (above 30~eV) 
analytically with a $1/\omega^2$ fall off.

Based on the mechanisms described above, carrier injection essentially leads to a change in the refractive index. 
Another significant contribution to the refractive index change is caused by the plasma absorption 
of the free carriers: electrons in the conduction band for the $n$-doped case
or the electrons and holes for the photoexcited case. Its resultant contribution can be expressed 
in convenient units as\cite{bennett}
\begin{equation}
\label{nplasma}
\Delta n_{\mbox{\begin{scriptsize}{plasma}\end{scriptsize}}}=-\frac{6.9\times 10^{-22}}
{n_0 m_r^*}\frac{N[\mbox{cm}^{-3}]}{E^2[\mbox{eV}]}\, ,
\end{equation}
where $n_0$ is the refractive index of the intrinsic semiconductor, 
$E[\mbox{eV}]$ is the energy in eV, $N[\mbox{cm}^{-3}]$ is the free-carrier density in cm$^{-3}$ 
and $m^*_r$ is the effective mass in units of $m_0$; in the photoexcited case, for the latter two 
we use the electron-hole density and the effective mass ($\mu_{eh}$), respectively.

\section{Results}
\subsection{$n$-doped case in InN and GaN}
\begin{figure}[h]
\begin{center}
\includegraphics[width=9cm]{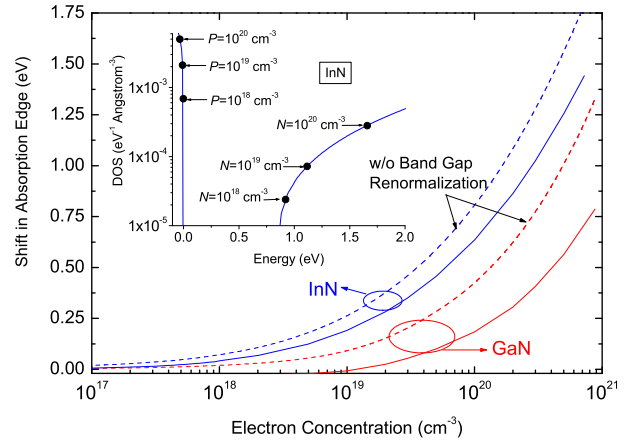}
\caption{\label{fig5} The shift in the absorption edge by $n$-type doping for 
GaN and InN due to conduction band filling without (dashed) and with (solid) 
BGR accounted. The inset shows the density of states (DOS) 
for InN together with the band-filling levels on the valence and conduction 
bands in the absence of BGR.}
\end{center}
\end{figure}

We first start with the $n$-doped case which is quite easily achieved (sometimes even 
unintentionally) in InN and GaN and, we explore the filling of the conduction band 
by electrons fully ionized from their dopants. The electron concentrations up 
to about 7$\times 10^{20}$~cm$^{-3}$ (9$\times 10^{20}$~cm$^{-3}$) are considered
marking the threshold beyond which the next higher conduction band starts to be 
filled for InN (GaN).
Figure~\ref{fig5} shows the shift in the absorption edge due to Pauli blocking of 
the filled portions of the lowest conduction band. A marked contrast revealed with 
this plot is that the absorption edge shift in InN is much more significant compared to GaN.
The disparity between the two materials' conduction band edge density of state behavior 
plays a role here. While in InN the band-filling effect is quite strong, in the case of GaN this 
becomes smaller which is moreover effectively canceled by the BGR. 

\begin{figure}[h]
\begin{center}
\includegraphics[width=9cm]{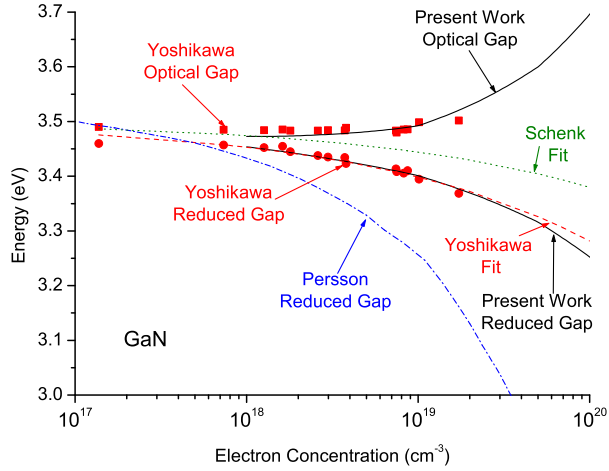}
\caption{\label{fig6} The comparison for the $n$-doped GaN, the reduced gap and the optical gap values 
(see text) with Yoshikawa {\em et al.},\cite{yoshikawa99} Schenk {\em et al.}\cite{schenk} and 
Persson {\em et al.}\cite{persson} The band gap of intrinsic GaN is taken as 3.5~eV.}
\end{center}
\end{figure}

\begin{figure}[h]
\begin{center}
\includegraphics[width=9cm]{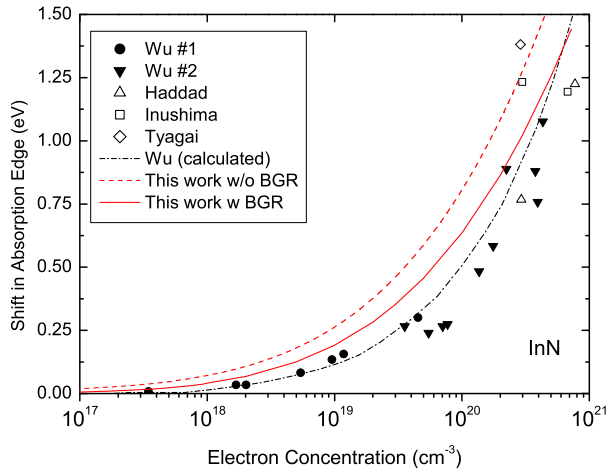}
\caption{\label{fig7} The comparison of measured and calculated shifts in the 
absorption edge in InN with the calculation and measurements from Wu,\cite{wu04} 
together with other data mentioned in that work: 
Haddad,\cite{haddad} Inushima,\cite{inushima} Tyagai.\cite{tyagai}}
\end{center}
\end{figure}

Yoshikawa {\em et al.} have used temperature-dependent photoluminescence spectroscopy 
to separate BGR and band-filling effects in Si doped GaN.\cite{yoshikawa99}
Figure~\ref{fig6} illustrates the comparison of our $n$-doped GaN results with this work for the 
reduced band gap, i.e., due to BGR, as well as for the optical gap which 
corresponds to the absorption edge including the BGR and 
Burstein-Moss effects. The dashed line is the fitted expression, $-4.72\times 10^{-8}n^{1/3}$~eV 
offered by Yoshikawa {\em et al.} for the former.
Very recently Schenk {\em et al.} relying on band edge luminescence peak and the line shape information have 
extracted an expression given by $-2.6\times 10^{-8}n^{1/3}$~eV for the BGR in 
$n$-doped GaN.\cite{schenk} 
While our result closely agrees with the Yoshikawa data, Schenk's expression is in disagreement with these. 
In Fig.~\ref{fig6}, also a much more elaborate theoretical analysis for the 
BGR is shown which however deviates from the rest of these results.\cite{persson} 
We believe that further studies are required to reach to a consensus over the BGR 
expression for GaN. The situation is more harmonious for InN.
The comparison of our computed absorption shift for InN with the measurements and calculation from 
Wu {\em et al.},\cite{wu04} together with other data mentioned in that 
work\cite{tyagai,inushima,haddad} is provided in Fig.~\ref{fig7}; it shows the 
overall agreement of our results with the literature. 

\begin{figure}[h]
\begin{center}
\includegraphics[width=9cm]{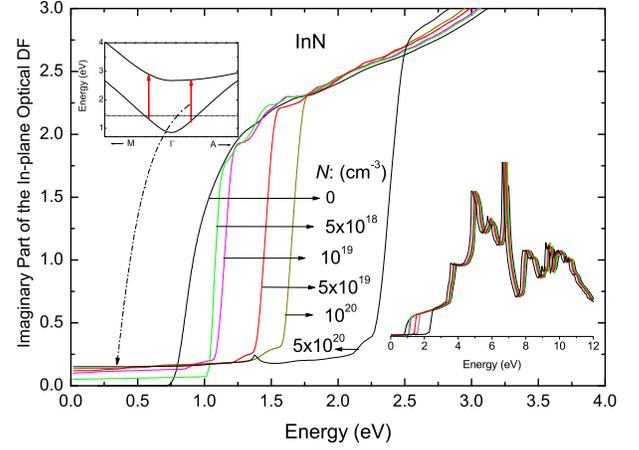}
\caption{\label{fig8} The calculated imaginary part of the in-plane dielectric 
function (DF) of InN for different $n$-type doping densities. Lower inset shows 
the same graph for a wider energy scale. Upper inset indicates the responsible 
intraconduction band transitions for the low energy part 
of the spectra.}
\end{center}
\end{figure}

\begin{figure}[h]
\begin{center}
\includegraphics[width=9cm]{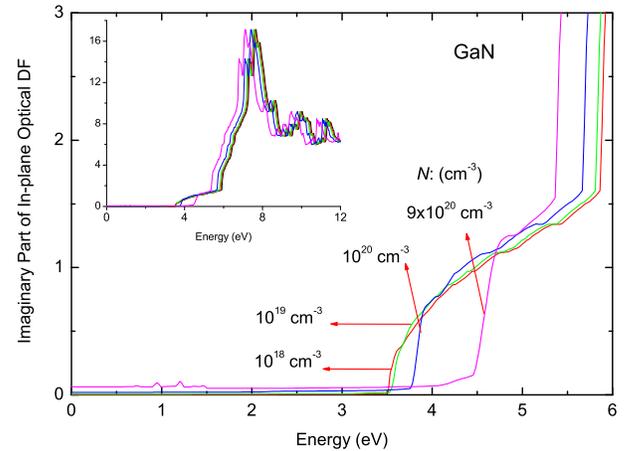}
\caption{\label{fig9} The calculated imaginary part of the in-plane dielectric function (DF) 
of GaN for different $n$-type doping densities. 
Inset shows the same graph for a wider energy scale.}
\end{center}
\end{figure}

The conduction band filling and the associated BGR directly 
affect the imaginary part of the dielectric functions through the shift in 
the absorption edge deeper into the conduction band.
This is illustrated for the ordinary components (i.e., the electric field lying on the 
crystal basal plane) for different electron concentrations in Figs.~\ref{fig8} and 
\ref{fig9} for InN and GaN, respectively. 
It can be noticed for both cases, especially for 
larger dopings, that there is a nonzero absorption below the absorption 
edge, even down to zero energy. This is due to {\em intraband} absorption within 
the conduction band (see the upper inset in Fig.~\ref{fig8}).
The same behavior also appears in the measured InN samples.\cite{wu04}

\begin{figure}[h]
\begin{center}
\includegraphics[width=9cm]{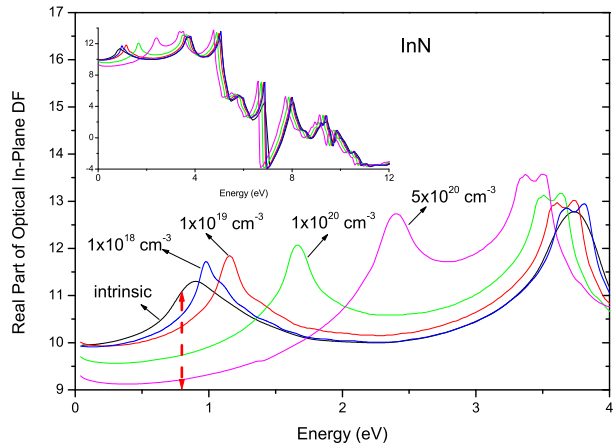}
\caption{\label{fig10} The calculated real part of the in-plane electronic dielectric function (DF) 
of InN for different $n$-type doping densities. The vertical red arrow marks the 0.8~eV (1.55~$\mu$m) value.
Inset shows the same graph for a wider energy scale.}
\end{center}
\end{figure}

\begin{figure}[h]
\begin{center}
\includegraphics[width=9cm]{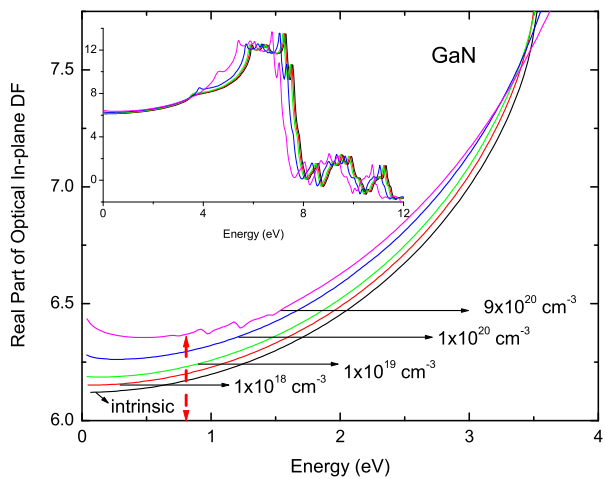}
\caption{\label{fig11} The calculated real part of the in-plane electronic dielectric function (DF) 
of GaN for different $n$-type doping densities. The vertical red arrow marks the 0.8~eV (1.55~$\mu$m) value.
Inset shows the same graph for a wider energy scale.}
\end{center}
\end{figure}

\begin{figure}[h]
\begin{center}
\includegraphics[width=9cm]{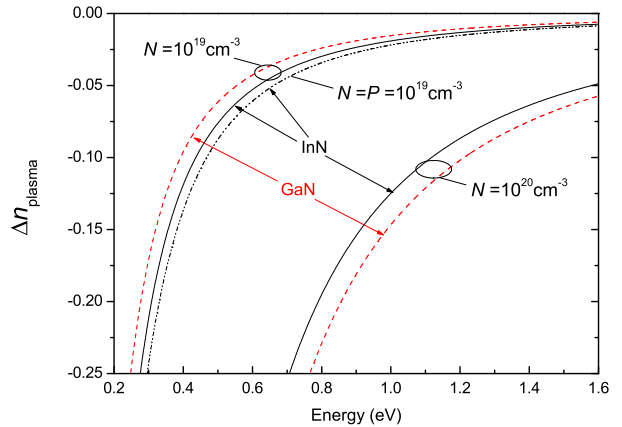}
\caption{\label{fig12} The comparison of the free-carrier-induced refractive index change, 
i.e., plasma contribution, [see, Eq.~(\ref{nplasma})] for two different $n$-type doping 
densities for InN and GaN. The photoexcited $N=P=10^{19}$~cm$^{-3}$ bipolar case in 
InN is also included.}
\end{center}
\end{figure}

The real parts of the dielectric functions are obtained from the imaginary parts using 
the Kramers-Kronig relation in Eq.~(\ref{kkr}).
The results are shown in Figs.~\ref{fig10} and \ref{fig11} for InN and GaN, respectively. 
First of all, the real part of dielectric function of InN is more 
sensitive to $n$-type doping. Furthermore, for low energies, the shifts are in opposite 
directions for InN and GaN, so that as the electron concentration increases, the 
permittivity decreases for InN, whereas it increases for GaN. 
Note that these are only the electronic contributions to polarization as Eq.~(\ref{im_eps}) 
does not include the ionic degrees of freedom. Nevertheless the ions do not directly play a 
role in the {\em change} in the refractive index which is focus of this work.
As mentioned in the Theory section, one should also add the free carrier 
plasma contribution to the refractive index [see, Eq.~(\ref{nplasma})]. 
This is separately shown in Fig.~\ref{fig12} which always has a negative contribution, and 
$|\Delta n_{\mbox{\begin{scriptsize}{plasma}\end{scriptsize}}}|$ 
decreases as $1/E^2$ with energy $E$. 

\begin{figure}[h]
\begin{center}
\includegraphics[width=9cm]{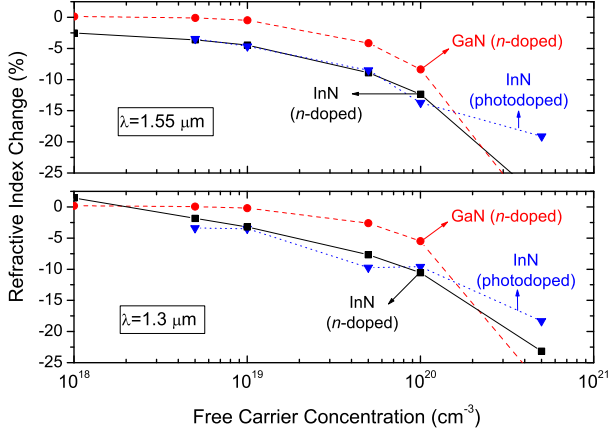}
\caption{\label{fig13} The percentage refractive index change with respect to intrinsic 
(i.e., no free carrier) case for InN and GaN. All effects considered in this work are included, 
i.e., band-filling, BGR, and plasma contributions.
Top panel shows the 1.55~$\mu$m wavelength (0.8~eV) and the bottom panel 
corresponds to the 1.3~$\mu$m wavelength (0.954~eV). See text, for the explanation of why 
$n$-doped InN curve in the bottom panel starts from a positive value. 
The lines are to guide the eyes.}
\end{center}
\end{figure}

The overall change in the refractive index, 
including all of the effects mentioned so far is shown in Fig.~\ref{fig13}. 
Because of their technological importance, it is evaluated at the 
two fiber-optic communication wavelengths of 1.55 and 1.3~$\mu$m which correspond to 
energies of 0.8 and 0.954~eV, respectively; the former 
is marked with dashed vertical arrows in Figs.~\ref{fig10} and \ref{fig11}.
It can be observed that the refractive index change due to $n$-type doping 
is much higher in InN compared to GaN for both wavelengths.  
There are several reasons behind this outcome. First of all there is a cancellation that occurs 
for GaN; as shown in Fig.~\ref{fig5}, BGR very effectively 
cancels the Burstein-Moss shift up to a density of 10$^{19}$~cm$^{-3}$. The 
lack of such a cancellation in InN is because of the
very strong band-filling effect as a result of small density of states of InN 
close to conduction band edge, see the inset in Fig.~\ref{fig5}.
Furthermore, contrary to the case in InN, the refractive index change in GaN 
due to band filling is 
positive up to the ultraviolet. This is canceled by the negative plasma 
contribution [cf., Eq.~(\ref{nplasma})], which actually dominates beyond a 
density of 10$^{19}$~cm$^{-3}$ in GaN and 10$^{20}$~cm$^{-3}$ in InN.

Note that at the 1.3~$\mu$m wavelength the intrinsic as well as doped InN 
display substantial interband absorption (see, Fig.~\ref{fig8}), 
hence, in Fig.~\ref{fig13} we used the following generalization of the refractive index in the 
presence of loss:\cite{klingshirn}
\begin{equation}
n=\sqrt{\frac{\sqrt{\mbox{Re}\{\epsilon\}^2+\mbox{Im}\{\epsilon\}^2}+
\mbox{Re}\{\epsilon\}}{2}} \, .
\end{equation}
Because of this loss, 1.3~$\mu$m case shows an overall reduction 
in the refractive index change compared to 1.55~$\mu$m wavelength. As another curious point, 
the $n$-doped InN curve in the bottom panel of Fig.~\ref{fig13} starts from a positive value 
which requires an explanation.
Due to the absorption in the 1.3~$\mu$m wavelength, the refractive index also peaks 
around the 0.95~eV range, particularly for the 10$^{18}$~cm$^{-3}$ density (cf. Fig.~\ref{fig10}). 
Hence, the $n$-doped InN curve in Fig.~\ref{fig13} becomes positive for 
this density as its refractive index {\em exceeds} that of the intrinsic sample. Even though 
the band-filling and plasma contributions try to reduce it, they are not as effective 
at this density.

\subsection{Photoexcited case in InN}
\begin{figure}[h]
\begin{center}
\includegraphics[width=9cm]{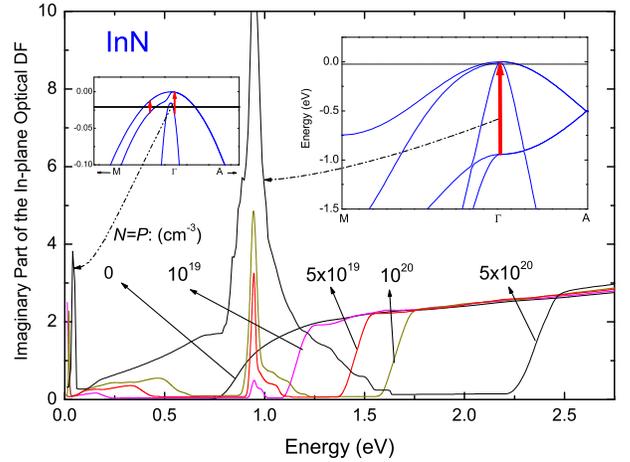}
\caption{\label{fig14} The calculated imaginary part of the in-plane dielectric function
of InN for different electron-hole populations forming degenerate quasiequilibrium distributions.
Insets show the intravalence band transitions that give rise to the new absorption channels 
around 0~eV (left) and 0.94~eV (right).}
\end{center}
\end{figure}

We would like to compare the pronounced refractive index change in $n$-doped InN with
the case where both electrons and holes are present. Since introducing holes with $p$-type 
doping is still a major technical challenge in InN,\cite{wu09} we consider 
photodoping\cite{photodoping} where equal density of 
electrons and holes are generated by a strong photoexcitation. 
We further assume that these photocarriers quickly thermalize to their respective band 
edges and attain a degenerate quasiequilibrium distribution.
Since the electron-hole pairs will eventually recombine within a time scale of 
300-400~ps,\cite{chen03} the quasicontinuous optical excitation should replenish this in order 
to sustain the assumed steady-state carrier distribution. 
The associated quasi-Fermi levels for different electron-hole densities are marked in 
the inset of Fig.~\ref{fig5} which clearly indicates that the contribution of the holes to the 
Burstein-Moss effect is negligible. In Fig.~\ref{fig14} we show the imaginary dielectric 
function at several electron-hole densities. Compared to the $n$-doped case in Fig.~\ref{fig8}, 
we observe two new features around 0 and 0.94~eV which are caused by the intravalence band transitions:
the former is due to very low energy transitions at the valence band maximum (left inset in 
Fig.~\ref{fig14}), whereas the latter is the result of the $\Gamma^v_5 \to \Gamma^v_6$ transition 
(right inset in Fig.~\ref{fig14}). It is the latter that makes a significant impact around 
1.55~$\mu$m (0.8~eV) and even more so at the 1.3~$\mu$m wavelength (0.954~eV). 
Therefore, we would like to assert that our band structure agrees quite well with 
the other first-principle studies\cite{bagayoko05,furthmuller} which report the value 
0.9~eV with the quasiparticle correction for the $\Gamma^v_5 \to \Gamma^v_6$ transition. 
As observed from Fig.~\ref{fig14}, these new 
intravalence band channels refill the part of the spectra cleared by the Pauli blocking, undoing 
the Burstein-Moss effect. 
Also we can observe from Fig.~\ref{fig13} that below a carrier density of 10$^{20}$~cm$^{-3}$, 
the change in the refractive index at 1.55~$\mu$m is seen to be very close to the $n$-doped case.
However, it should be recalled that this is now accompanied by a strong absorption (cf., 
Fig.~\ref{fig14}) which can hamper its possible device applications. 

\section{A self-critique of the theoretical model and discussions}
In this section we would like to discuss several aspects of the employed theoretical model 
both for the assertion of its general validity as well 
as for pointing out its possible improvements. 
To start with, all our results follow from the dielectric 
function expression of Eq.~(\ref{im_eps}) which is obtained 
within the independent particle (i.e., random phase) approximation,\cite{ehrenreich} 
and as such, it neglects the excitonic effects. In the case of intrinsic InN 
dielectric function, it has been shown by Furthm\"uller {\em et al.} that, apart from 
small shifts in the energy positions, these excitonic effects are responsible 
for the enhancement of some of the peaks in the spectra, predominantly close to the 
band edge, while the general shape remaining as unaffected.\cite{furthmuller} 
Furthermore, in the presence of high concentration of free carriers, as considered 
in this work, the excitonic effects get perfectly screened. The transition from 
bound excitonic state to free electron-hole plasma takes place at the 
Mott critical density.\cite{klingshirn} If we apply its estimate from 
Ref.~\onlinecite{bennett}, we get about 5$\times10^{16}$ and 5$\times10^{18}$~cm$^{-3}$ 
for InN and GaN, respectively. Above these values, excitons will dissociate which is 
easily met for the density range of this study.

On the BGR, there are other theoretical estimations, 
such as the plasmon-pole\cite{haug} and the Vashishta-Kalia\cite{vk} models. 
However, we observed that when contrasted with the existing experimental data on 
GaN\cite{nagai} and InN,\cite{schley} the employed Berggren-Sernelius expression\cite{berggren} 
in this and recent studies\cite{wu02b,wu04} outperforms these other models in the 
relevant high density range of this study. From a formal point of view, 
there exist more rigorous many-body approaches which incorporate 
excitons, BGR and band-filling effects in the same level, 
however, their formidable computational cost leads to a trade off with 
a simplistic effective mass band structure.\cite{lowenau}

Furthermore, we should mention that the employed Berggren-Sernelius expression assumes a 
cubic crystal with parabolic bands.\cite{berggren} These simplifications have been avoided by 
Persson and co-workers who have incorporated the electron-phonon contribution as well.\cite{persson}
Their framework is particularly better suited for handling the BGR under 
photoexcitation where both electron- and hole-gas many-body contributions can be individually 
included. However, as we stated previously, in the photoexcited case we still use 
the Berggren-Sernelius expression but with the reduced electron-hole effective mass. 
Essentially, the same discussion also applies for the plasma contribution in Eq.~(\ref{nplasma}).

Compared to the available {\em ab initio} results for 
InN,\cite{rinke,furthmuller,carrier05,bagayoko05} our band structure is in good 
overall agreement. Nevertheless, there remain further discrepancies of most 
of these theoretical results from the experimental data. 
First of all, due to the fact that our computational band gap of 0.85~eV for InN is 
somewhat higher than the experimentally established value,\cite{wu09} we expect 
especially our 1.55~$\mu$m wavelength data to be affected. However, this change should 
be more or less similar to the deviation between 1.55~$\mu$m and 1.3~$\mu$m 
cases in Fig.~\ref{fig13}. In the case of GaN, our observation of a substantial 
deviation from the measured absorption in the vicinity of band gap 
(see, Fig.~\ref{fig3}) calls for improvement on pseudopotential form factors for 
GaN which were originally optimized for high-field transport applications.\cite{prb00} 
Furthermore, we do not include the band tail (Urbach's tail) absorption from the impurity 
states forming bands below the band edge.\cite{shen02,arnaudov} Just like the excitons, 
the effect of localized states become marginal for high free-carrier densities, like beyond 
10$^{18}$~cm$^{-3}$. Moreover, the characteristic 
Urbach energy parameter in high-quality InN samples turn out to be as low as 
7.5 to 15~meV which indicates that these features are anchored to just 
below the band gap.\cite{klochikhin}

Being intended as a bulk study, this work neglects any surface effects, the most important 
of which is the
surface charge accumulation causing built-in electric field and band bending.\cite{lu03,mahboob}
In this regard, it has been mentioned by Kudrawiec {\em et al.} that for $n$-type InN layer since 
the region of strong band bending is the surface electron accumulation layer, it will not affect 
absorptionlike techniques such as photoreflectance spectroscopy due to band filling 
of the conduction band.\cite{kudrawiec}
Another effect discarded in this study is the thermal smearing of the degenerate carrier 
statistics. As a possible follow up of this work it can be readily included, nevertheless 
we do not expect any marked deviations. 
Even though the quasi-Fermi level shift in the valence band is of the same order as the 
thermal energy (cf., inset of Fig.~\ref{fig5}), as shown above, the main effect is driven 
by the Fermi level shift in the conduction band, which by far outweighs any thermal 
energy well up to room temperature for the carrier densities important for this work.
Stemming from these idealizations, our main results in Fig.~\ref{fig13} should be taken as 
{\em upper bounds}.
The list of these simplifications also suggests possible improvements of this work.

To harness the carrier-induced refractive index change such as in phase modulators, 
the associated losses need to be minimized.
In this respect, the narrow 0.64~eV band gap of InN becomes undesirable for the 1.55 and 
1.3~$\mu$m applications as they enable supra band gap excitations where additional 
losses set in, especially for the photoexcited case. Our comparative study here suggests 
that the In$_x$Ga$_{1-x}$N can enjoy both the high carrier-induced refractive index 
sensitivity while at the same time operating in the transparency region of the alloy. 
The poor sensitivity of GaN to carrier-induced refractive index change, as demonstrated 
by this work points towards the choice of indium-rich alloys.

\section{Conclusions}
We have shown that based on a full band theoretical analysis, 
under $n$-type doping, InN and GaN should display different dielectric characteristics. 
The absorption edge shift is masked in GaN, whereas it is highly pronounced in InN. 
For energies below 1~eV, the corresponding  shifts in the real parts of the 
dielectric function for InN and GaN are in opposite directions. The free-carrier plasma 
contribution to refractive index becomes dominant above 10$^{20}$~cm$^{-3}$ 
for the case of InN. At a wavelength of 1.55~$\mu$m, we predict more than 4\% of 
change in the refractive index for InN for a doping of 10$^{19}$~cm$^{-3}$. 
Under optical pumping which fills the conduction and valence band edges with electrons and holes  
of equal density, the refractive index is not further affected by the presence of the holes. 
However, its major consequence is the increased intravalence band aborption as a result of 
$\Gamma^v_5 \to \Gamma^v_6$ transition. Beyond a hole density of 10$^{19}$~cm$^{-3}$, this affects 
1.3 to 1.55~$\mu$m wavelengths.
We believe that this wide tunability of the index of refraction in InN 
by the conduction band electrons is technologically important for applications 
such as optical phase modulators. In the presence of holes, 
the valence band absorption needs to be taken into account, such as in the design of InN-based 
lasers for fiber-optics applications.
Our findings suggest that the alloy composition of In$_x$Ga$_{1-x}$N can be optimized in the indium-rich 
region so as to benefit from high carrier-induced refractive index change while operating in the 
transparency region to minimize the losses.

\begin{acknowledgments}
The authors would like to thank the British Council for supporting the partnership of the Essex and 
Bilkent Universities. CB and CMT would like to thank the EU FP7 Project UNAM-Regpot Grant No. 203953 
for partial support.
\end{acknowledgments}

\begin{table}[h]
\caption{Fitted values for the constants in the pseudopotential form factors
 $V_s$ and $V_a$; see Eqs.~(1) and (2). Other parameters used in the EPM are
 also listed, where $a$ and $c$ are the lattice constants in the hexagonal
 plane and along the $c$ axis, respectively, $u$ is the wurtzite internal structural
 parameter, and $E_{\text{max}}$ denotes the radius of the energy sphere
 used for the reciprocal lattice vectors.}
\vspace*{0.5cm}
\begin{tabular}{l l}
\hline\hline
$s_1=0.261905$ & $a_1=-0.411320$ \\
$s_2=0.232300$ & $a_2=0.176416$ \\
$s_3=-2.726784$ & $a_3=1.441420$ \\
$s_4=0.973335$ & $a_4=0.915637$ \\
$s_5=2.414422$ & $a_5=2.595997$ \\ \hline
$a=3.544\,$\AA & \\
$c=5.716472\,$\AA & \\
$u=0.379$ & \\
$E_{\text{max}}=12.5\,$Ry\\
\hline\hline
\end{tabular}
\end{table}
\begin{table}[h]
\caption{Fitted nonlocal EPM parameters for InN. $A$ values are in Rydbergs and $R$ values are in Angstroms.}
\vspace*{0.5cm}
\begin{tabular}{l l}
\hline\hline
$A^{\mbox{\begin{scriptsize}In\end{scriptsize}}}_p$=5.55440, & $R^{\mbox{\begin{scriptsize}In\end{scriptsize}}}_p$=0.195168\\
$A^{\mbox{\begin{scriptsize}In\end{scriptsize}}}_d$=4.51525, & $R^{\mbox{\begin{scriptsize}In\end{scriptsize}}}_d$=0.132069\\
$A^{\mbox{\begin{scriptsize}N\end{scriptsize}}}_p$=3.64897, & $R^{\mbox{\begin{scriptsize}N\end{scriptsize}}}_p$=0.089251\\
\hline\hline
\end{tabular}
\end{table}
\appendix*
\section*{Appendix: Empirical Pseudopotential Parameters for wurtzite I\lowercase{n}N}
In the light of recent experimental and first-principles studies on wurtzite InN, we have 
expressed the {\em local} empirical pseudopotential method (EPM) form factors using the following 
functional form,
\begin{eqnarray}
\label{Vs}
V_s(q) & = & \left( s_1 q^3+s_2 q^2+s_3\right)\,\exp{(-s_4\/ q^{s_5})}\,
 , \\
\label{Va}
V_a(q) & = & \left( a_1 q^2+a_2 q+a_3\right)\,\exp{}(-a_4\/ q^{a_5})\, ,
\end{eqnarray}
where $V_s$ and $V_a$ are the symmetric and antisymmetric form
 factors in Rydbergs, $q$ is the wave number in units of
 $2\/\pi/a$ with $a$ being the lattice constant in the hexagonal plane,
 and $s_i$, $a_i$ $(i=1,\ldots ,5)$ are the fitting parameters which
 are listed in Table~I.

The {\em nonlocal} part of the pseudopotential is based on the classical work of Chelikowsky and Cohen,\cite{cc}
with the plane-wave matrix elements being given by
\begin{eqnarray}
\label{VNL}
V_{NL}(\vec{K},\vec{K'}) & = & \frac{4\pi}{\Omega_a}\sum_{l,i}A^i_l(E)(2l+1)P_l\left(\cos(\theta_{KK'})\right)
\nonumber \\ & & \times 
S^i(\vec{K}-\vec{K}')F^i_l(K,K'),
\end{eqnarray}
where $S^i(q)$ is the structure factor\cite{cc} for the atomic species $i$, and $P_l(x)$ is the Legendre polynomial 
corresponding to angular-momentum channel $l$, $\Omega_a$ is the atomic volume (i.e., primitive cell volume divided by 
number of atoms), $\vec{K}=\vec{k}+\vec{G}$, $\vec{K}'=\vec{k}+\vec{G}'$, and 
\begin{widetext}
\begin{equation}
F_l(K,K')=\left\{
\begin{array}{ll}
\frac{R^3}{2}\left\{ \left[j_l(KR)\right]^2-j_{l-1}(KR)j_{l+1}(KR)\right\}, & K=K',  \\
\frac{R^2}{K^2-K'^2}\left[K j_{l+1}(KR)j_{l}(K'R)-K'j_{l+1}(K'R)j_{l}(KR)\right], & K \neq K'
\end{array}
\right.
\end{equation}
here, $j_l(x)$ is the spherical Bessel function, $R$ is (angular-momentum-dependent) well width. 
For InN, we treat the angular-momentum 
channels $p$ and $d$ for In, and $p$ for N as nonlocal. Table~II lists InN nonlocal EPM parameters.
\end{widetext}


\begin{thebibliography}{99}
\bibitem{photodoping}There exists a somewhat different use of the 
``photodoping'' terminology to describe the dissolution of  metals, such as Ag or Cu, 
into chalcogenide glasses under an illumination. However, this is more appropriately 
termed as photodissolution.\cite{kolobov}
\end{thebibliography}
\end{document}